\documentclass[lettersize,journal]{IEEEtran}
\usepackage{amsmath,amsfonts}
\usepackage{algorithmic}
\usepackage{algorithm}
\usepackage{array}
\usepackage[caption=false,font=normalsize,labelfont=sf,textfont=sf]{subfig}
\usepackage{textcomp}
\usepackage{stfloats}
\usepackage{url}
\usepackage{verbatim}
\usepackage{graphicx}
\usepackage{cite}
\usepackage{multicol} 

\begin{document}

\title{Ten issues of NetGPT}

\author{
	\begin{flushleft}
		\IEEEauthorblockN{
			Wen Tong$^{1}$, Chenghui Peng$^{1}$, Tingting Yang$^{2}$, Fei Wang$^{1}$, Juan Deng$^{3}$, Rongpeng Li$^{4}$, Lu Yang$^{1}$, Honggang Zhang$^{5}$, Dong Wang$^{6}$, Ming AI$^{7}$,  Li Yang$^{8}$, Guangyi Liu$^{3}$,  Yang Yang$^{9}$, Yao Xiao$^{1}$,
			Liexiang Yue$^{3}$, Wanfei Sun$^{7}$, Zexu Li$^{6}$, Wenwen Sun$^{8}$
		}
		\begin{multicols}{2}

			$^1$Wireless Technology Lab, Huawei Technologies\\
			$^2$Pengcheng Lab\\
			$^3$China Mobile Research Institute\\
			$^4$Zhejiang University\\
			$^5$Zhejiang Lab\\

			$^6$Research Institute of China Telecom\\
			$^7$CICT Mobile Communication Technology Co., Ltd.\\
			$^8$ZTE Corporation\\
			$^9$The Hong Kong University of Science and Technology (Guangzhou)
		\end{multicols}
	\end{flushleft}
}
\maketitle

\begin{abstract}
	With the rapid development and application of foundation models (FMs), it is foreseeable that FMs will play an important role in future wireless communications. As current Artificial Intelligence (AI) algorithms applied in wireless networks are dedicated models that aim for different neural network architectures and objectives, drawbacks in aspects of generality, performance gain, management, collaboration, etc. need to be conquered. In this paper, we define NetGPT (Network Generative Pre-trained Transformer) - the foundation models for wireless communications, and summarize ten issues regarding design and application of NetGPT.
\end{abstract}


\section{Introduction}
\subsection{Background}
\IEEEPARstart{T}{he} integration of artificial intelligence (AI) and communication has been proposed as one of the most important usage scenarios of International Mobile Telecommunications (IMT) for 2030 and beyond (IMT-2030) by International Telecommunication Union (ITU) \cite{ITU}. Many people have been devoted to this research field over the past few years and fruitful study results are achieved, which can benefit various aspects of mobile networks from operation, administration and management (OAM), Core Network (CN), Radio Access Network (RAN) to UE (User Equipment). For example, deep deterministic policy gradient (DDPG) has been utilized in policy generation of CN; deep Q-learning (DQN) has been used for network OAM, etc. However, such existing paradigm of the tailored AI algorithm for each specific use case will lead to many problems, such as low generality, limited performance gain, complicated management, and inconvenient collaboration \cite{tong2021challenges}\cite{letaief2019roadmap}\cite{yang20226g}.

As a breakthrough development of AI techniques, foundation models have been fascinatedly applied to many fields. The most well-known one is large language model (LLM), which has exhibited its powerful capabilities in chatting, programming, etc. It can be expected that FMs with a relatively unified neural network architecture will advance the wireless network in terms of high generality, superior performance gain, simplified management, convenient collaboration, multi-task processing capability, etc. To distinguish from other industrial applications, we name FMs for wireless communications as NetGPT. Nevertheless, the design and application of NetGPT is still at an initial stage. In this paper, we summarize ten fundamental issues of NetGPT to advance the design of both the model and the wireless network architecture for supporting NetGPT application efficiently.

\subsection{Definition of NetGPT}
Wireless communication network is comprised of different technical domains like RAN/CN/OAM, which vary tremendously in function features, data structures, performance requirements, etc. Huge differences are anticipated when deploying NetGPT in each part of a wireless network. As an example, NetGPT can be applied to OAM by fine tuning LLMs via Parameter-Efficient Fine-Tuning. The second type of NetGPT can be obtained by distilling or pruning LLMs, for edge deployment with relatively smaller size. The third type of NetGPT can be trained from scratch, with similar neural network architecture of LLMs. Therefore, NetGPT is not a single model accounting for all scenarios of wireless communications, but a series of models covering different technical domains and vendors.

\begin{figure*}[!t]
	\centering
	\includegraphics[width=6in]{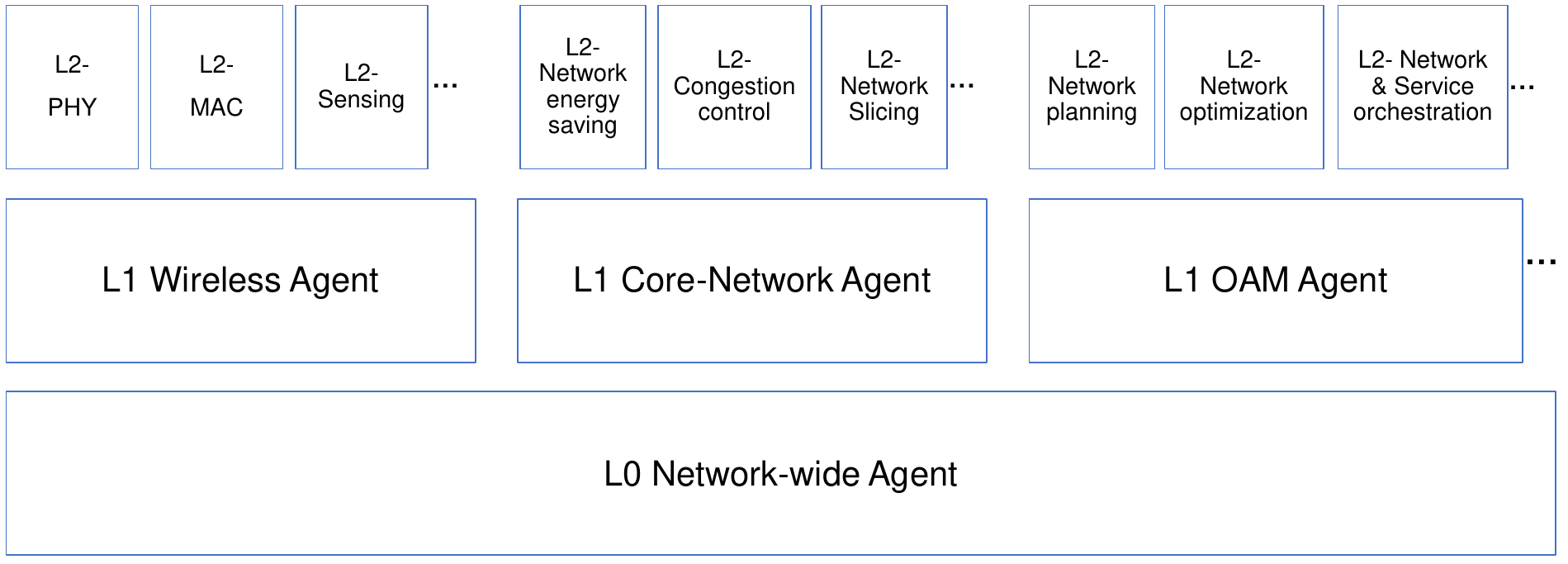}
	\caption{Three layers of agents for NetGPT}
	\label{fig_1}
\end{figure*}

In this paper, we define three layers of agents for NetGPT, namely, layer 0 (L0), layer 1 (L1) and layer 2 (L2). NetGPT-L0 agent represents large network-wide model. NetGPT-L1 agent refers to the FMs for different technical domains, such as RAN, CN, or OAM. NetGPT-L2 agent indicates the models are focused on more specialized scenarios, for example, NetGPT-L2 agent for physical layer of RAN and NetGPT-L2 agent for network optimization of OAM, as shown in Fig.\ref{fig_1}.

\section{Ten Issues of NetGPT}
In this section, we propose ten issues of NetGPT regarding wireless communications for 6G. These issues can be classified into two categories. One category focuses on the design of NetGPT itself, whereas the other attends to the design of future wireless network architecture for supporting the application of NetGPT, including RAN, CN, and OAM system.

\subsection*{Issue 1: Scenarios and Requirements of NetGPT}
Trained on enormous datasets and billions of parameters, NetGPT demands high computing power. Unlike the centralized and powerful computing capability of cloud, that of wireless networks comprising multiple edge devices and mobile terminals is heterogeneous and distributed. Therefore, instead of a direct deployment of cloud AI models/algorithms to wireless networks, NetGPT requires major redesigns. Specifically, it involves adaptation to the characteristics of wireless networks in algorithm design, and native support of NetGPT in network architecture design. In addition, it is worth noting that the lower the layer in a wireless network, the higher the requirement of quality of service (QoS) including real-time and accuracy. The complications like hallucination and extra-large size of LLMs hinder the fulfillment of the QoS.  A key problem thus arises: whether there exists a boundary of applying NetGPT in wireless networks. For example, whether NetGPT is applicable only to the higher layer of the air-interface, but not the physical layer? Such boundary issues also engage the potential role that NetGPT can play in each specific application. For example, to what extent can NetGPT enable future OAM systems - fully or partial autonomous networks \cite{tmf}.
It is necessary to clarify the basic scenarios and boundary issues mentioned above for study of NetGPT.

\subsection*{Issue 2: Theoretical Gaps between NetGPT and LLM}
As the most representative FM, it has been pointed out that LLM could form a basis for NetGPT \cite{chen2023netgpt}\cite{bariah2023large}\cite{shen2023large}.  However, there are many essential differences between communication field and natural language processing (NLP) field, which leads to theoretical gaps between NetGPT and LLM. The primary distinctions are as follow:

\begin{itemize}
	\item{Data Features: The data corpus of NetGPT contains communication information, for instance, the channel information. Such information is represented in a form of high-dimensional tensors rather than tokens in the LLM.}
	\item{Back-end task: Wireless networks deal with different types of tasks. The output for NetGPT consequently may exist in different forms, in lieu to tokens that serve as both the input and output of LLMs.}
	\item{Model size: NetGPT defines a hierarchy, with models of various sizes deployed at each level. Some NetGPT models, especially the NetGPT-L2 deployed at network edges (e.g. BSs), may have only 0.1$\sim$1B parameters, far less than 5$\sim$200B parameters contained in typical centralized LLMs.}
\end{itemize}

The following problems await further investigation: whether the neural network architecture of NetGPT can be consistent with that of LLM; whether NetGPT triggers evolutionary innovations in AI theories and the associated neural network architecture.

\begin{figure}[!t]
	\centering
	\includegraphics[width=3.5in]{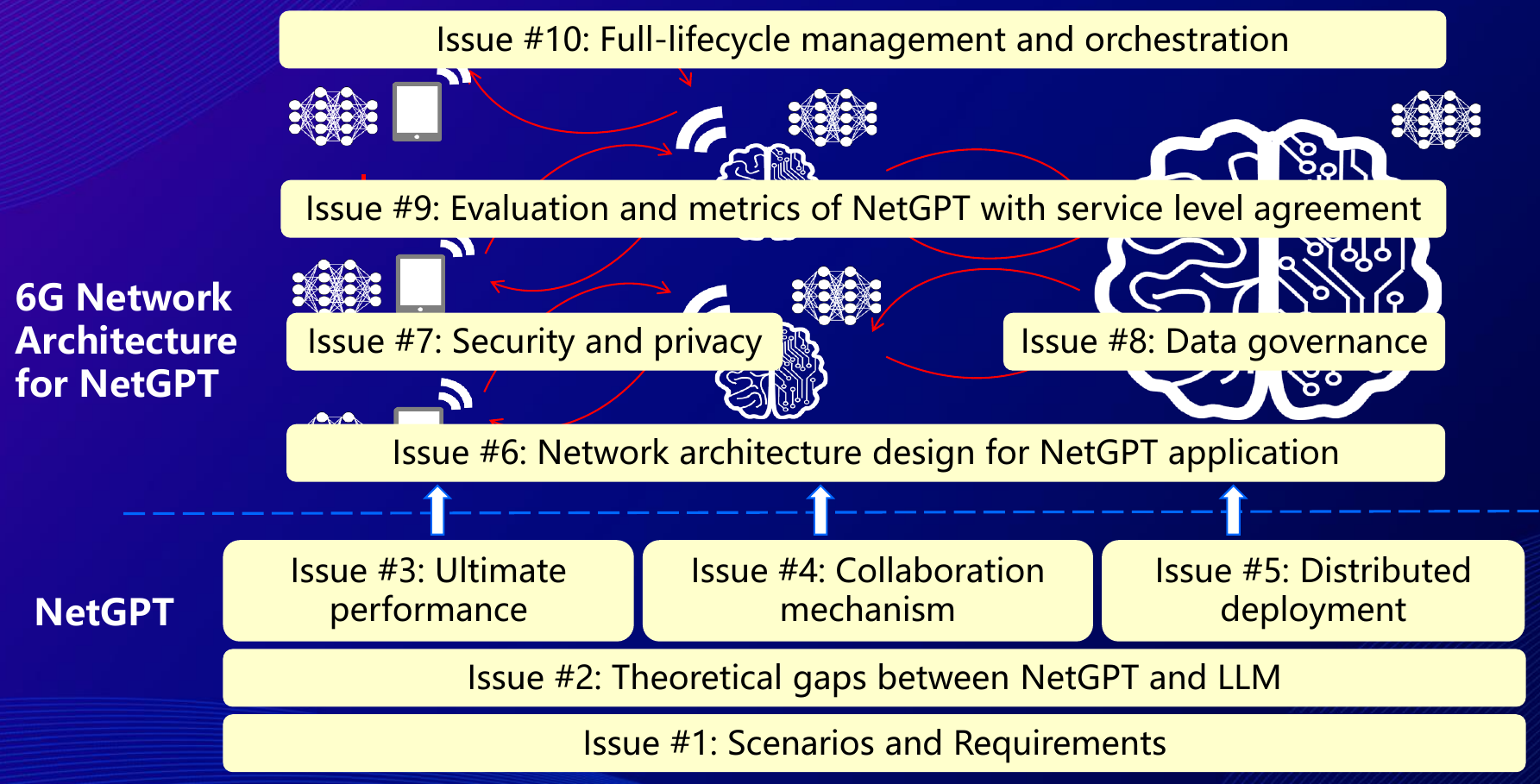}
	\caption{Ten Issues of NetGPT}
	\label{fig_2}
\end{figure}

\subsection*{Issue 3: Ultimate Performance of NetGPT}
The requisite for applying NetGPT to wireless networks is to fulfill the ultimate performance requirements (e.g. rapid real-time and reliability) of future wireless networks such as 6G. These requirements of future wireless networks will be much higher than the achievable ones by current FM applications.

Compared to 5G, future wireless networks need to not only boost 10x improvement in the communication performance, but also support the new services assured by such ultimate performance, like the autonomous driving, the industrial robots, etc.  Integration of AI and communication allows NetGPT to achieve the aforementioned goals for future wireless networks. That is, NetGPT needs to derive inference results in a very short time, and adapt to the highly dynamic environment of wireless networks in real time. Due to complicated computing procedures and a huge number of parameters, it is very challenging for NetGPT to achieve 0.1ms-level real-time inference in future wireless networks. More efficient model algorithms and inference acceleration methods are therefore desired.

In addition, future wireless networks will have extremely high requirements on the reliability \cite{ITU}. Since hallucination problem of FMs may cause incorrect network decisions and unpredictable risks, most of current FMs play an auxiliary role, lacking a direct application to network service functionalities. To meet the high reliability requirements of future wireless networks, methods to enhance the reliability of NetGPT should be explored from the perspectives of data quality, model structure and knowledge graph, etc.

\subsection*{Issue 4: Collaboration Mechanism of NetGPT}
Considering extravagant computation, communication and energy resources that the FMs may cost, models with smaller size like NetGPT-L2 may better suit the edge of wireless network like BSs. While NetGPT-L2 is advantageous to deal with specific scenarios with local knowledge, it may lack necessary generality under certain circumstances. This hitch could be resolved by the collaboration with central-deployed NetGPT-L0/L1.

Thus, the deployment of NetGPT in future wireless network will emphasize the collaboration between NetGPT with various scales, which involves collaborative training and inference. In particular, collaborative training scenarios include: (1) NetGPT-L0/L1 integrates local data to elaborate locally tailored NetGPT-L2 via transforming, distillation, pruning, etc; (2) NetGPT-L0/L1 aids the training process of NetGPT-L2 as an auxiliary factor; (3) feedback from NetGPT-L2 further boosts NetGPT-L0/L1. On the other hand, collaborative inference scenarios include: (1) Real-time inference is accomplished collaboratively with NetGPT-L0/L1 when NetGPT-L2 is not able to reach the desired confidence level independently; (2) Real-time inference is accomplished collaboratively by multiple NetGPT-L2 at each terminal. In above scenarios, certain key algorithms await further investigation, such as parameter pruning according to relevance from NetGPT-L0 parameters, and efficient fine-tuning to accommodate new tasks.

In addition to the algorithmic challenges, to support cross-vendor collaboration of NetGPT, a standardized collaboration mechanism including functionalities and procedures need to be defined, such as standardization of collaboration content, generation methods of collaboration set, and systematic control of collaborative incidents.

\subsection*{Issue 5: Distributed Deployment of NetGPT}
In terms of end-edge-cloud collaboration, it might require to deploy, at the wireless network edge or terminal, a full or partial version of NetGPT-L0, L1 or third-party FMs, due to service requirements.  We need to exploit how to split FMs to better accommodate to the heterogeneity of wireless network end-edge-cloud and dynamic environment. As each node has different computing resources, storage capabilities and transmission rates, the actual availabilities should determine the split to achieve optimal load balancing as well as resource utilization.

Cross node incremental training for consistency at wireless edge also needs to be exploited. The parameter update of distributed subnets of a FM may bear a lag time, which causes inconsistency. A mechanism is needed to guarantee parameter consistency for incremental training of FMs in the network.

Moreover, an efficient communication mechanism among distributed nodes is another core factor. One approach is to reduce communication overhead by algorithm optimization like model compression (e.g. pruning and quantization). Another one is to enhance the efficiency of data flow among nodes by refinement of interface and protocols.

\subsection*{Issue 6: Network Architecture Design for NetGPT Application}
With growing integration of NetGPT integrated into the future wireless network, NetGPT will bring evolutionary impact on the following perspectives of network architectures:

For network element, NetGPT is capable of complete or partial realization of existing network functions. However, as NetGPT being widely applied to end-to-end scenario, certain classic network functionalities will be outdated and reorganized/redesigned thoroughly, such as network element classification for current 5G core network, along with the way to organize them (e.g. service-based architecture) \cite{6gana1}.

For interface and the corresponding protocol, model-based collaborative interface (e.g. tokenized interface) among NetGPT will substitute classic interface protocols built upon standardized signals and element strings.

For new network capacities, the future wireless network natively supports NetGPT and third-party FMs, including online update and evolution of various kinds of NetGPT. Different from the incremental training via PEFT (e.g. low-rank adaptation, LoRA) for accommodating FMs to specific scenarios, the kind of incremental training considered in NetGPT usually requires learning of new knowledge in addition to the general abilities of FMs. A life-long learning technique is thus desired. On the other hand, as data parallel (e.g. Federated Learning) and model parallel (e.g. Split Learning ) may be inadequate for NetGPT due to immense number of parameters and data, the emergence of both is worth exploring.

For new network services, based on the ability of FMs to interpret semantic information, the future wireless network can generate an exclusive network for each individual application, providing services accordingly from perspectives such as business logics, network logics and resources.

\subsection*{Issue 7: Security and Privacy of NetGPT}
When NetGPT is applied in end-to-end wireless networks, the security of NetGPT will be highly concerned. The involvement of huge amount of data and parameters in FMs yields attack vulnerability of the NetGPT, particularly conscious implant of backdoors, which might confuse NetGPT with illegal instructions. Moreover, NetGPT may become more biased and untrustworthy as they become larger. All of these issues require tailored security design for NetGPT.

As the interpretability of NetGPT is inheritly unavailable like other AI algorithms, its application in network is at risks. Although some important research theories have been proposed, it lacks a well-established theoretical framework, especially regarding the mathematical and analytical approaches for quantitative analysis of FMs.

In addition, during the use of NetGPT, users' privacy data, such as user account information, dialog history, and information obtained during interactions, are possibly exposed to suppliers, service providers, and affiliates. Global regulatory agencies also have begun to pay attention to the data privacy risks brought by FMs after multiple data leakage incidents. Therefore, data privacy principles and usage regulations for NetGPT are desperately desired in applications.

\subsection*{Issue 8: Data Governance of NetGPT}
As the performance of NetGPT highly depends on data quality, the future wireless network requires data governance service specially designed for NetGPT \cite{6gana2}:

Firstly, future wireless network needs to support efficient processing of massive heterogeneous data. Data fed into NetGPT comes from various technical fields and parties, which demands a comprehensive framework for data governance in terms of data proprietary rights, formats, qualities, privacy, etc.

Secondly, future wireless network needs to have large-scale data distributed storage and real-time provision mechanism for inference and online updating of NetGPT. Especially for the NetGPT with ultimate performance requirements, how to design corresponding data service with guaranteed QoS will be a challenge.

Thirdly, future wireless network needs to establish network knowledge graph for boosting the reliability of NetGPT and reducing hallucination.

Based on the above new data governance capabilities, a unified data governance framework needs to be designed in the future wireless network for different types of NetGPT.

\subsection*{Issue 9: Evaluation and Metrics of NetGPT with Service Level Agreement}
How to evaluate the NetGPT comprehensively and objectively has become an urgent problem to be solved. On the one hand, the evaluation of the performance of models can form a strong basis for the optimization and improvement of NetGPT, and thus improve its application effect and commercial value. On the other hand, the evaluation can be used as a benchmark to understand the performance and applicability of NetGPT provided by different vendors.

Besides existing evaluation methods, such as accuracy, F1 Score \cite{Rijsbergen}, Bilingual Evaluation Understudy \cite{Bleu} and so on, a set of evaluation indicators based on network characteristics should be formulated, such as network function correctness, communication task success rate. The combination of these specific indicators helps to evaluate the performance of NetGPT in specific scenarios in a more refined manner. And for in-network tasks, we still need to pay attention to the understanding and application of domain-specific terms, concepts, and rules to ensure the reliability of evaluation results.

Different from the language domain, network is relatively closed and fewer annotation data could be collected publicly. The difficulty of collecting the small sample data causes an incomprehensive scope of possibly encountered scenarios in the model training. Therefore, the generalization of NetGPT to different network scenarios should also be evaluated.

\subsection*{Issue 10: Full-lifecycle Management and Orchestration of NetGPT}
NetGPT from different vendors are deployed across the network. A unified mechanism needs to be established within the network to provide full lifecycle orchestration and management of NetGPT, such as adding, updating, transferring, and removing. In particular, it is important to ensure that the intellectual property rights of NetGPT are well reserved during this process. As the owners of NetGPT are not necessarily the operator of the network, they may not be willing to allow the network operator control their models. Therefore, a well-balanced scheme is yearned for protecting the interests of both sides.

Another challenge is how to organize and schedule these NetGPT. First, we need to reconcile language for models from different vendors, that is, consistent interfaces and interaction methods. Then we must consider that the deployment of NetGPT involves multi-dimensional resources such as connections, computation, and storage. Accordingly, it will be a delicate task to orchestrate the models and network resources based on scenario characteristics and the requirements to improve system performance and resource utilization.

\section{Conclusion and Prospect}
Albeit the potential revolutionary change might be introduced by FMs to the future wireless networks, many related research directions and standardization of FMs demand further investigations. NetGPT implies a bidirectional deep convergence between wireless networks and FMs. In this paper, we proposed ten fundamental issues of NetGPT, including basic theories, scenario requirements, network architecture, deployment management and control, and data governance. We call for further in-depth research of NetGPT to push forward the integration of FMs in future wireless networks.


\begin{thebibliography}{10}

	\bibitem{ITU}
	ITU-R, ``Framework and overall objectives of the future development of {IMT}
	for 2030 and beyond,'' 2023.

	\bibitem{tong2021challenges}
	W.~Tong and G.~Y. Li, ``Nine challenges in artificial intelligence and wireless
	communications for {6G},'' {\em IEEE Wireless Communications}, vol.~29,
	no.~4, pp.~140--145, 2022.

	\bibitem{letaief2019roadmap}
	K.~B. Letaief, W.~Chen, Y.~Shi, J.~Zhang, and Y.-J.~A. Zhang, ``{The Roadmap to
	6G: AI Empowered Wireless Networks},'' {\em IEEE Communications Magazine},
	vol.~57, no.~8, pp.~84--90, 2019.

	\bibitem{yang20226g}
	Y.~Yang, M.~Ma, and H.~Wu, ``{6G Network AI Architecture for Everyone-Centric
	Customized Services},'' {\em IEEE Network}, pp.~1--10, 2022.

	\bibitem{tmf}
	T.~McElligott, ``Network automation using machine learning and {AI},'' {\em
	TMForum}, 2020.

	\bibitem{chen2023netgpt}
	Y.~{Chen}, R.~{Li}, Z.~{Zhao}, C.~{Peng}, J.~{Wu}, E.~{Hossain}, and
	H.~{Zhang}, ``{{NetGPT: A Native-AI} Network Architecture Beyond Provisioning
	Personalized Generative Services},'' {\em arXiv e-prints},
	p.~arXiv:2307.06148, July 2023.

	\bibitem{bariah2023large}
	L.~{Bariah}, Q.~{Zhao}, H.~{Zou}, Y.~{Tian}, F.~{Bader}, and M.~{Debbah},
	``{Large Language Models for Telecom: The Next Big Thing?},'' {\em arXiv
	e-prints}, p.~arXiv:2306.10249, June 2023.

	\bibitem{shen2023large}
	Y.~{Shen}, J.~{Shao}, X.~{Zhang}, Z.~{Lin}, H.~{Pan}, D.~{Li}, J.~{Zhang}, and
	K.~B. {Letaief}, ``{Large Language Models Empowered Autonomous Edge {AI} for
	Connected Intelligence},'' {\em arXiv e-prints}, p.~arXiv:2307.02779, July
	2023.

	\bibitem{6gana1}
	6GANA, ``{Ten Questions of 6G Native AI Network Architecture},'' {\em
	https://www.6g-ana.com/}, 2022.

	\bibitem{6gana2}
	6GANA, ``{6G Data Service - Concept and Requirements},'' {\em
	https://www.6g-ana.com/}, 2022.

	\bibitem{Rijsbergen}
	C.~J. Van~Rijsbergen, ``Information retrieval (2nd ed.),'' {\em
	Butterworth-Heinemann.}, 1979.

	\bibitem{Bleu}
	K.~Papineni, S.~Roukos, and T.~Ward, ``{BLEU}: a method for automatic
	evaluation of machine translation[c],'' {\em Proceedings of the 40th annual
	meeting of the Association for Computational Linguistics.311-318}, 2002.

\end{thebibliography}
\end{document}